\documentclass[letterpaper,twocolumn,10pt]{article}
\usepackage{endnotes}

\usepackage{authblk}

\usepackage[T1]{fontenc}
\usepackage[english]{babel}

\usepackage{booktabs}

\usepackage{url}

\usepackage[colorlinks=true]{hyperref}

\usepackage{subfigure}
\usepackage[pdftex]{graphicx}
\DeclareGraphicsExtensions{.pdf,.gif}

\usepackage{textcomp} %

\usepackage{amssymb}
\usepackage{amstext}
\usepackage{amsmath}

\usepackage{todonotes}
\usepackage{lmodern, textcomp}
\usepackage{paralist}

\begin{document}

\title{Squeezing out the Cloud via\\Profit-Maximizing Resource Allocation Policies}

\author{Michele Mazzucco}
\author{Martti Vasar}
\author{Marlon Dumas}
\affil{\it University of Tartu\\Estonia}
\maketitle

\begin{abstract}
We study the problem of maximizing the average hourly profit earned by a Software-as-a-Service (SaaS) provider who runs a software service on behalf of a customer using servers rented from an Infrastructure-as-a-Service (IaaS) provider. The SaaS provider earns a fee per successful transaction and incurs costs proportional to the number of server-hours it uses. A number of resource allocation policies for this or similar problems have been proposed in previous work. However, to the best of our knowledge, these policies have not been comparatively evaluated in a cloud environment. This paper reports on an empirical evaluation of three policies using a replica of Wikipedia deployed on the Amazon EC2 cloud. Experimental results show that a policy based on a solution to an optimization problem derived from the SaaS provider's utility function outperforms well-known heuristics that have been proposed for similar problems. It is also shown that all three policies outperform a ``reactive'' allocation approach based on Amazon's auto-scaling feature.
\end{abstract}

\section{Introduction}
\label{sec:introduction}

Two key pillars of cloud computing are the notions of \emph{elasticity}, wherein resources are available in any amount and at any time, and \emph{pay-per-use}, wherein users are charged only for the resources they consume.
In an idealized conception of these notions, cloud computing consumers are able to acquire exactly the amount of computing resources they need and add or release resources instantly in order to cope with changes in workload. However, computing resources (specifically virtual servers) require some setup time ({\it e.g.},  5--10 minutes), resources are acquired in discrete units ({\it e.g.} one server of a given capacity), and billing is done for discrete time intervals ({\it e.g.}, one hour). As a result, cloud consumers must carefully balance the tradeoff between their cost-reduction objective, which pushes them to acquire a minimum number of servers, and the imperative of dealing with varying workloads, which requires them to keep some ``slack''.

In this setting, this paper considers the case of a Software-as-a-Service (SaaS) provider who runs a service on behalf of a customer using resources provided by an Infrastructure-as-a-Service (IaaS) provider. In line with the pay-per-use model, the customer pays a fixed charge to the SaaS provider per successful transaction, subject to a Service Level Agreement (SLA) specifying performance objectives. Meanwhile, the SaaS provider incurs a cost proportional to the number of server-hours used. 

As any economic actor, the SaaS provider seeks to maximize its \emph{profit}, that is, the total fees charged by the SaaS provider to its customer minus the cost of renting servers from the IaaS provider and, when applicable, the penalties paid by the SaaS provider for violating SLA objectives. Thus, a key question faced by the SaaS provider is how to maximize profit under varying workload and knowing that additional resources take some time to become operational and they are acquired for discrete time intervals. 

This question can be reduced to a utility-driven resource allocation problem, that is, the problem of determining the optimal amount of resources (servers in this case) to be allocated during a given epoch in order to maximize a given utility function (the profit function in this case). To address this problem, we consider three resource allocation policies aimed at maximizing the mean profit that the SaaS provider earns per hour. 
The first policy is obtained by solving an optimization problem derived from the utility function. The other two correspond to heuristics that have been proposed for similar problems in previous work.
To make them operational, the policies are coupled with a model to forecast the workload for the upcoming epoch and methods to estimate service rate and throughput under a given configuration. The operationalized policies are comparatively evaluated based on a replica of Wikipedia deployed on the Amazon cloud.

The paper is structured as follows. Section~\ref{sec:model} introduces the adopted system model and spells out the assumptions. Section~\ref{sec:estimation} presents the workload, service rate and throughput estimation methods. Section~\ref{sec:policies} introduces the resource allocation policies.
Section~\ref{sec:experiments} presents the experimental evaluation. Finally, Section~\ref{sec:relatedwork} discusses related work while Section~\ref{sec:conclusion} outlines directions for future work.

\section{System Model and Utility Function}
\label{sec:model}

We consider a SaaS provider that at any given point in time has $n$ virtual servers available to deal with incoming requests for a given service ({\it jobs}, from now on). The virtual servers are assumed to be homogeneous in terms of their performance. This assumption can be called into question for two reasons. Firstly, cloud providers such as Amazon EC2 offer different types of virtual servers ({\it e.g.}, small, large, XL) with different capacities. However, as far as these differences go, the assumption of homogeneity can be relaxed by means of a normalization function~\cite{rykov:04}. Secondly, previous studies on the Amazon EC2 cloud have out into evidence non-negligible differences in terms of CPU and I/O performance between virtual servers of the same type ({\it e.g.} small instances)~\cite{jiang:2011}.
Moreover, a given virtual server can behave differently over time, for example due to startup/shutdown of other virtual servers on the same physical server\endnote{\url{http://www.infoq.com/news/2010/01/ec2-oversubscribed}}. Such differences are unpredictable and ultimately, all we can assume is a certain ``minimum'' level of capacity per virtual server. Accordingly, for the purpose of constructing a system model, the worst case scenario is assumed, that is, a homogeneous cluster of servers, each providing a fixed (guaranteed) capacity that is determined empirically as discussed later.

Every processed job generates a fixed revenue, $c$. This revenue might for example come from advertisements or from sales (in case of online merchants, such as Amazon).
In any case, it is assumed that the $c$ is given.  An incoming job that finds all the servers busy is blocked and lost (no charge is received by the SaaS provider) without affecting future arrivals. 
For each running server, the SaaS provider pays a fee of \$$d$ per hour to an IaaS provider. The IaaS provider bills per server-hour, regardless of whether the server is used for an entire hour or part thereof.
Given the above, the average profit $P$ earned by the SaaS provider per unit time is:

\begin{equation}
P = cT - dn\mbox{,}
\label{eq:revenue}
\end{equation}

\noindent where $T$ is the system's throughput.

The problem of the SaaS provider is to determine how many servers $n$ to acquire during a given time period. This problem is addressed by means of a \emph{resource allocation policy} that seeks to optimize the profit. The resource allocation policy is invoked periodically and each time it is invoked, it returns a value of $n$. The SaaS provider acquires the number of servers calculated by the policy. This implies acquiring additional servers or releasing some servers. We assume that data is replicated, hence releasing servers does not affect service availability.

During the intervals between consecutive policy invocations, the number of running servers remains constant. Those intervals, herein referred to as `epochs', are used to collect traffic statistics for the next policy invocation.

Since lost jobs do not generate revenue, the policy needs to balance the tradeoff between service availability and $n$. In the extreme scenario where, on aggregate, the charge per job is smaller than the cost paid for running the job, it is preferable not to use any server. Conversely, if the charge per job is orders of magnitude higher than the cost paid for running the job, the policy should provision the system for the expected peak workload so as to serve all jobs. The challenge is to design a policy that adapts to all scenarios in-between these two extremes.

The proposed system model does not distinguish cached and uncached jobs. A cached job is treated as any job since it requires a database access at least and it generates revenue as other jobs.

\noindent {\bf Extensions}
One could envisage alternative utility functions. For example, one might want to introduce a penalty for denial of service, $s$. In that case, Eq.~\eqref{eq:revenue} would be $cT - dn - (\lambda - T) s$ where $\lambda$ is the arrival rate, and thus $(\lambda - T)$ represents the rate at which traffic is rejected. As we shall see later, it is perhaps worth stressing that one can discourage denial of service even with Eq.~\eqref{eq:revenue} by properly tuning the value of $c$. Also, if the provider incurs a cost of $c_{1}$ for adding a server and a cost of $c_{2}$ for releasing a node, then the objective function would become $cT - dn - c_{1} n^{+} - c_{2} n^{-}$, where $n^{+}$ and $n^{-}$ represent the number of added and removed servers. None of those changes would alter the analysis, although they might alter the optimal solution. In the rest of this paper we will focus on utility function~\eqref{eq:revenue}.

\section{Throughput Estimation}
\label{sec:estimation}

The only unknown in function~\eqref{eq:revenue} is the throughput, which in turns necessitates the solution of a queueing model. %
No assumption is made about the nature of the arrival and service time processes; therefore, for a certain traffic intensity $\rho = \lambda / \mu$, with $\lambda$ and $\mu$ being the arrival and service rate respectively, and for a fixed number of servers $n$ ({\it i.e.}, the number of virtual machines), we model the number of jobs inside the system as a $G/GI/n/n$ queue.
Due to the fact that excess traffic is discarded, queues with finite buffer are always stable. It is therefore important to estimate how often clients are expected to experience a denial of service. 
Since no exact solution exists for the $G/GI/n/n$ queue we employ Hayward's approximation ({\it e.g.}, see~\cite{whitt:1984a}) to estimate the blocking probability $p_{n}$, as follows:

\begin{equation}
p_{n} = B\left(\frac{n}{z}, \frac{\rho}{z}\right) \mbox{,}
\label{eq:blocking}
\end{equation}

\noindent where $B(\cdot)$ stands for the Erlang-B formula~\cite{mitrani:1998} and $z$ is the asymptotic peakedness of the arrival process, {\it i.e.}, the variance divided by the mean of the steady-state number of busy servers in a $G/GI/\infty$ queue with the same arrival and service rate processes. The peakedness factor can be estimated as

\begin{equation}
z = 1 + (ca^{2} - 1) \eta {\mbox ,}
\label{eq:erlangB_peakedness}
\end{equation}

\noindent where $ca^{2}$ is the squared coefficient of variation ({\it i.e.}, variance over the square of the mean) of the interarrival intervals and $\eta$ is defined as

\begin{equation}
\eta = \mu \int_{0}^{\infty} [1 - G(t)]^{2} dt {\mbox ,}
\label{eq:erlangB_eta}
\end{equation}

\noindent and $G(t)$ is the cumulative distribution function (CDF) of the service time distribution with  mean $1 / \mu$ and variance $\sigma_{s}^{2}$. When $ca^{2} = 1$ the arrival process is Poisson, and thus Equation~\eqref{eq:blocking} is exact. When $ca^{2} \neq 1$ and the service times are exponentially distributed $z$ is equal to $1 + (ca^{2} - 1) / 2$, while when both the interarrival intervals and service times follow a general distribution, we 
approximate the distribution of $G(t)$ as the distribution of $N(1/\mu, \sigma_{s}^{2})$.

Having estimated the blocking probability, the average number of jobs entering the system (and completing service) per unit time is

\begin{equation}
T = \lambda [1 - p_{n}] \mbox{.}
\label{eq:throughput}
\end{equation}

\begin{figure}[hbt]
\centering
\subfigure[]{
\includegraphics[width=0.4\textwidth]{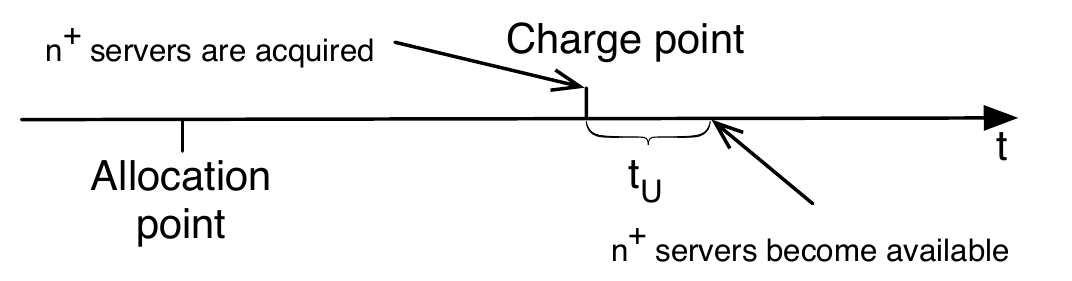}
\label{fig:allocation_up}
}
\subfigure[]{
\includegraphics[width=0.4\textwidth]{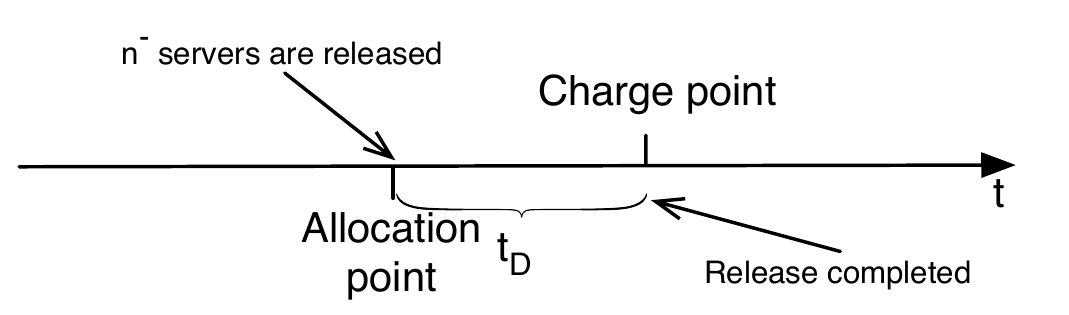}
\label{fig:allocation_down}
}
\caption{(a) Adding and (b) removing servers.}
\label{fig:allocation_up_down}
\end{figure}

If adding/releasing servers is not instantaneous, the formula for estimating the throughput is more complicated. Suppose that each configuration interval lasts $k$ time units ({\it e.g.}, 60 minutes), while bootstrapping new servers requires on average $t_{U}$ unit times (this includes not only acquiring the VM from the IaaS provider, but also operations such as updating the configuration of the infrastructure, synchronizing the state of the system, etc.). The additional $n^{+}$ servers are added at the billing instant (see Figure~\ref{fig:allocation_up}), while the average throughput for the {\it next} configuration interval can be estimated as

\begin{equation}
T^{+} = \frac{t_{U}}{k} \lambda [1 - B(n, \rho)] + \frac{k-t_{U}}{k} \lambda [1 - B(n+ n^{+}, \rho)] \mbox{,}
\label{eq:throughput+}
\end{equation}

\noindent where the first part of the above expression is the throughput of the system during the bootstrap of the $n^{+}$ extra nodes, while the cost is $d (n - n^{+})$.

If powering down servers requires at most $t_{D}$ time units (including VM termination but also system configuration update), the allocation decision is taken $t_{D}$ time units before the charge instant and $n^{-}$ servers are removed immediately, see Figure~\ref{fig:allocation_down}. Thus the average throughput for the {\it current} configuration interval reduces to

\begin{equation}
T^{-} = \frac{k-t_{D}}{k} \lambda [1 - B(n, \rho)] + \frac{t_{D}}{k} \lambda [1 - B(n - n^{-}, \rho)] \mbox{,}
\label{eq:throughput-}
\end{equation}

\noindent where the second part of the equation is the throughput of the system while $n^{-}$ servers are being released.
The throughput for the {\it next} interval is simply $\lambda[1 - B(n - n^{-}, \rho)]$, while the cost is $d (n - n^{-})$.
Note that when releasing servers the worst-case scenario should be considered for $t_{D}$, otherwise one might be charged for one full extra hour.

\noindent {\bf Parameters Estimation}
\label{sec:param_estimation}

It is assumed that up to $m$ jobs can be processed in parallel on each server without significant interference~\cite{kamra:2004, schwartz:2010, gunther:2002}, with such a limit being dictated by the number of available threads or processes\endnote{\url{http://httpd.apache.org/docs/2.0/mod/prefork.html}}.
Accordingly, in this study we estimate the service rate as the average throughput achieved for a certain value of $m$. A number of experiments discussing the model calibration are reported in Section~\ref{sec:calibration}.

\begin{figure}[ht!]
\centering
\includegraphics[width=0.48\textwidth]{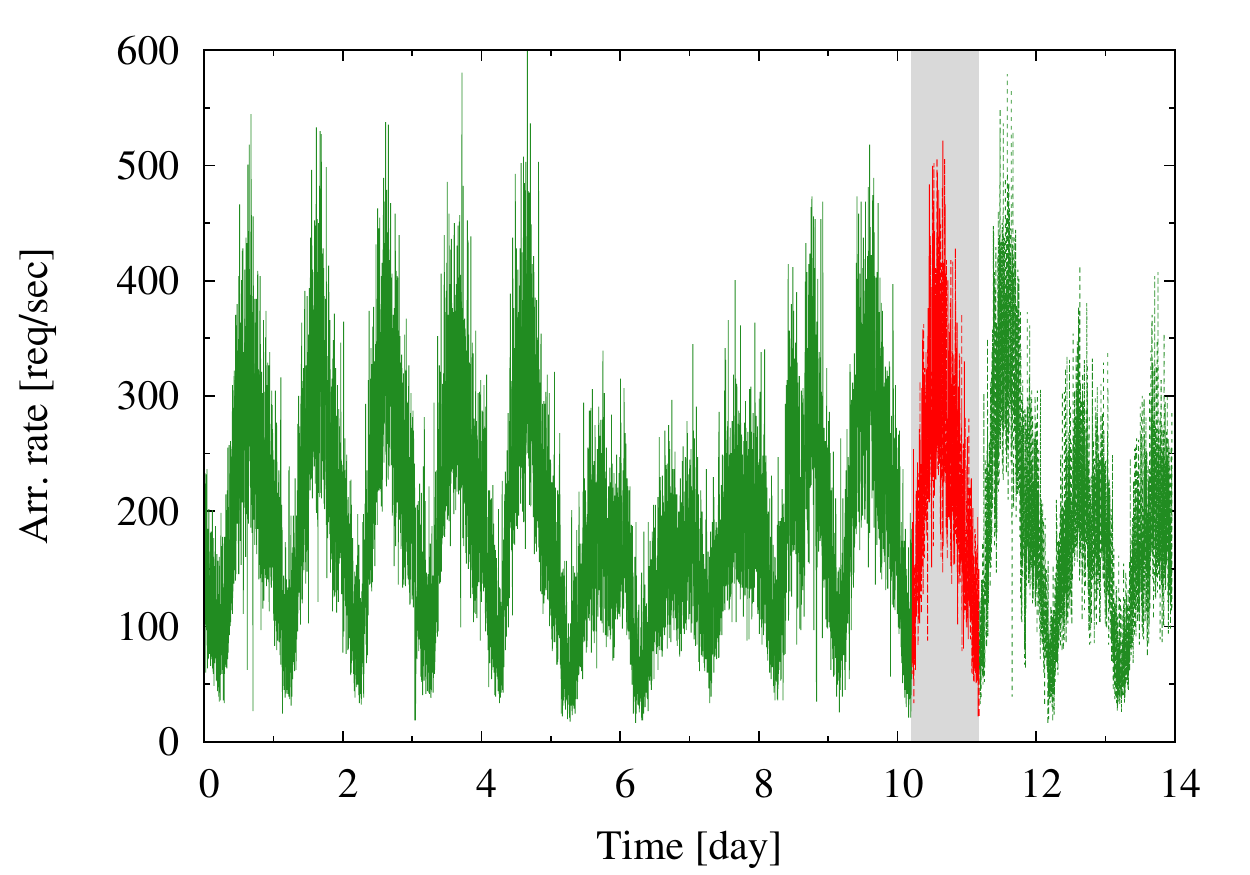}
\caption{ClarkNet workload (scaled version), one minute granularity. The red part (day 10) was used to evaluate the performance of our approach.}
\label{fig:clarknet}
\vspace{-1mm}
\end{figure}

The second parameter required to estimate the blocking probability is the arrival rate, $\lambda$. Unfortunately this value can be rarely estimated with absolute accuracy.
For example Figure~\ref{fig:clarknet} shows the arrival rate of the ClarkNet workload\endnote{\url{http://ita.ee.lbl.gov/html/contrib/ClarkNet-HTTP.html}} with one minute accuracy over a two weeks period. As one can see, the arrival rate exhibits a general trend, with daily and weekly patterns, as well as unexpected traffic spikes, which are hard to predict. An analysis of Wikipedia logs revealed similar patterns~\cite{mazzucco:2012}.

Different prediction algorithms exist to deal with time-varying user demand, which differ in complexity and accuracy~\cite{makridakis:2000, hyndman:2008}. In this study, we employ a modified Holt-Winters' algorithm with multiplicative effects for both seasonal and error components, specifically the ETS(M,A,M) algorithm~\cite{hyndman:2008} as implemented by the R command {\tt ets(ts, model="MAM", damped=F)}.

Finally, Equations~\eqref{eq:erlangB_peakedness} and~\eqref{eq:erlangB_eta} necessitate the squared coefficient of variations of interarrival intervals and service times. Those values can be easily estimated from the collected statistics. 

\section{Policies}
\label{sec:policies}

This section introduces three allocation policies, starting with a policy based on the solution of the utility function and moving on to profit-maximization heuristics proposed in previous work.

\subsection{Optimal Policy}
\label{sec:optimal_policy}

Once the parameters introduced in Section~\ref{sec:estimation} have been estimated the expressions~\eqref{eq:blocking} and~\eqref{eq:throughput} enable the utility function~\eqref{eq:revenue} to be computed efficiently. This allows us to determine the optimal number of servers to allocate, by computing the profit for different values of $n$ and finding the optimum value. When computing the utility function for different values of $n$, it becomes clear that $P$ is a unimodal function with respect to the number of servers, {\it i.e.}, it has a single maximum (Eq.~\eqref{eq:throughput} is convex for $n > 1$).

\noindent {\bf N.B.}  %
While the `Optimal' policy acknowledges the time $t_{U}$ necessary to launch new servers, it deliberately ignores the time $t_{D}$ required to terminate unnecessary nodes because the introduction of a short, but non-zero power-down interval has little effect on the `Optimal' policy. The rationale behind that decision is that servers are released when they are not needed anymore, so the reject probability of the last $t_{D}$ unit times of the current interval is not likely to be affected.

\begin{figure*}[ht!]
\centering
\subfigure[]{
\includegraphics[width=0.31\textwidth]{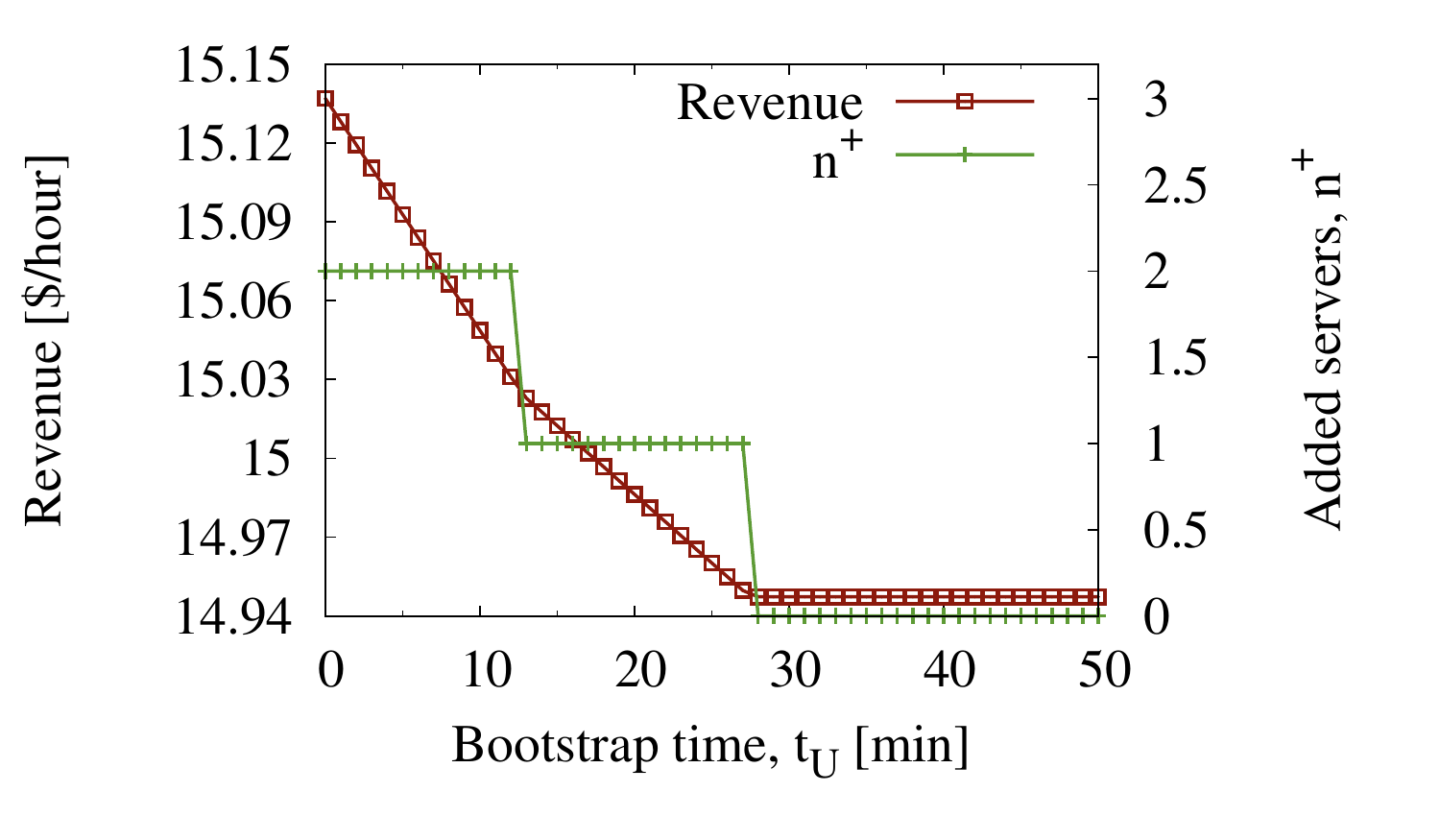}
\label{fig:R_tU}
}
\subfigure[]{
\includegraphics[width=0.31\textwidth]{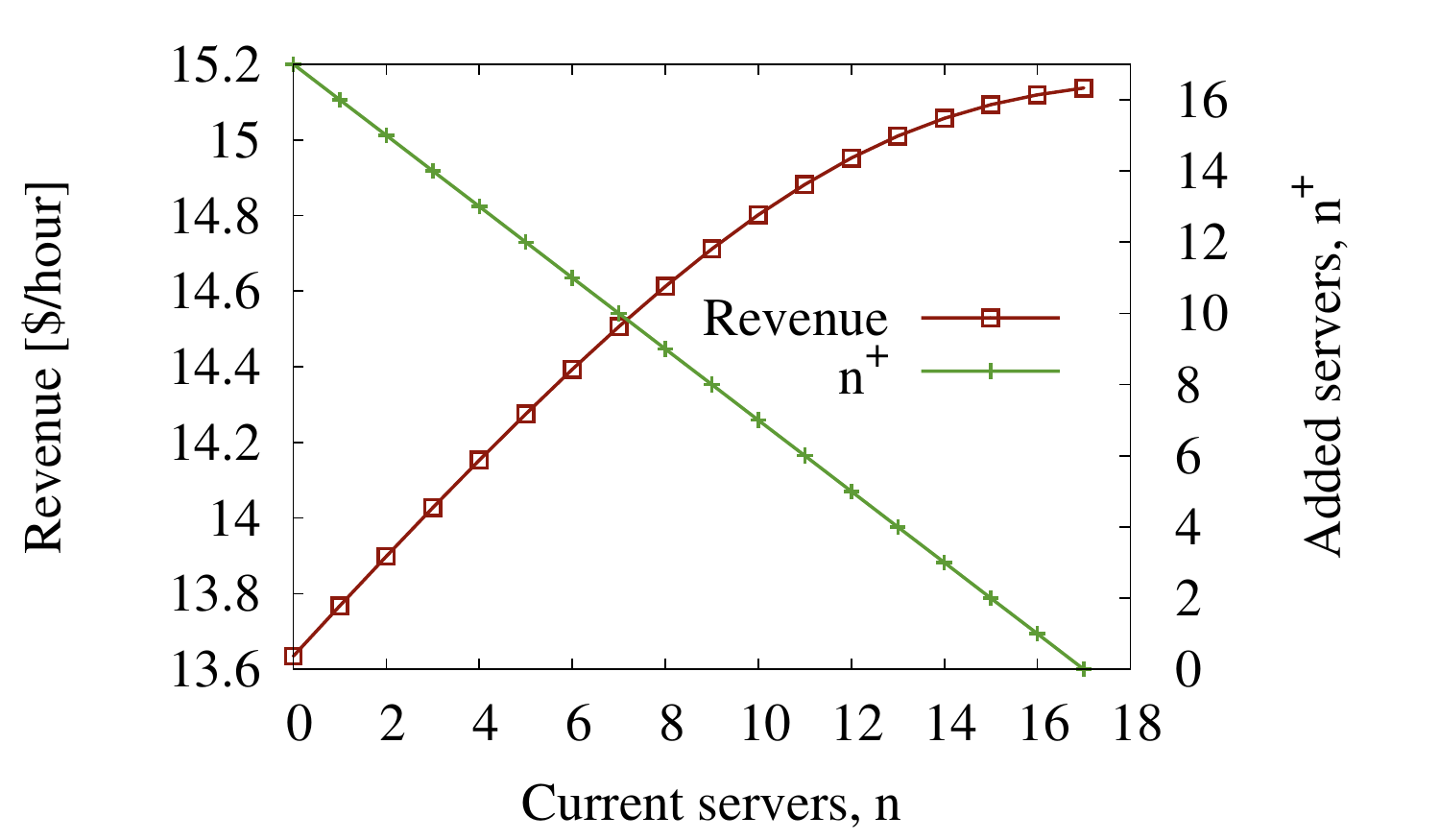}
\label{fig:R_n}
}
\subfigure[]{
\includegraphics[width=0.31\textwidth]{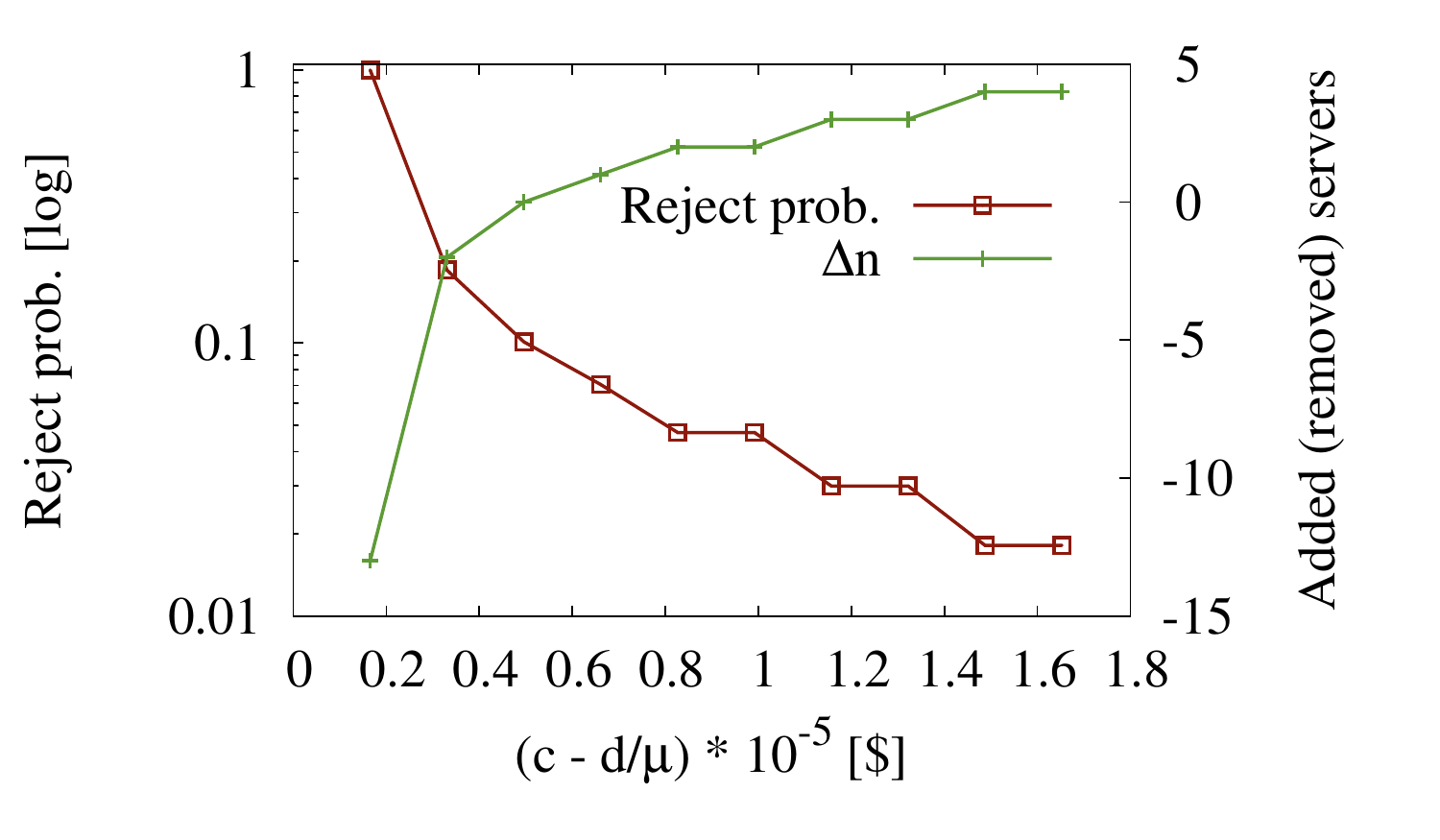}
\label{fig:dos}
}
\caption{(a) Revenue and added servers as a function $t_{U}$ and (b) revenue and added servers as a function $n$, and (c) reject probability and number of added (or removed) servers as a function of the difference between charge and cost per job. $\lambda = 300$, $k =1$ hour.}
\label{fig:sensitivity_analysis}
\end{figure*}

\paragraph{Sensitivity analysis}
Next, we assess how the `Optimal' policy reacts to changes in the interval required to acquire new servers, $t_{U}$, current number of servers, $n$, and charge minus cost per job, $(c - d/\mu)$.
In the following numerical experiments the arrival rate is fixed at $\lambda = 300$ jobs/sec., the service rate is $\mu = 28.571$ jobs/sec., while each configuration interval lasts one hour.

First we study how the optimal profit and the number of added servers, $n^{+}$, vary when $t_{U}$ increases from 0 to 50 minutes. In this experiment the number of servers running at the allocation point is held fix at $n = 15$.
As one might expect, an increase of $t_{U}$ results in a decrease of the number of extra servers as well as the profit. In fact, Figure~\ref{fig:R_tU} shows that $n^{+}$ decreases in a step-wise manner, while $P$ decreases in a linear manner.

Next, we study how the revenue and number of extra servers are affected by the value of $n$. Obviously, the throughput is maximized, and thus
the highest profit is achieved when no servers have to be added (this includes the scenario where some servers are released).
Figure~\ref{fig:R_n} also shows that the profit does not depend linearly on~$n^{+}$ (the cost does, but not the throughput).

Finally, we fix the bootstrap time to $t_{U} = 5$ minutes, $n = 13$, $d = 17$\textcent/hour and study how the value of $c$ affects the job loss and number of added/removed servers. As shown in Figure~\ref{fig:dos} when the difference between charge and cost per job, $(c - d/\mu)$, is lower than a certain threshold, the best strategy is to release all servers and reject all traffic. The value of this threshold depends on $\rho$ and $n$ as the reject probability is not constant with respect to the ratio $\rho/n$. In other words, if a $u$-fold increase in the offered load is matched by a concomitant $u$-fold increase of $n$, the reject probability decreases, {\it i.e.},  fewer servers (in proportion) are needed to guarantee the same service quality,
see Table~\ref{tab:non_constant}. %
On the other hand, an increase in the difference between charge and cost per job results in fewer servers being removed. Servers are added in larger measure as the charge is further increased, thus reducing the reject probability.

\begin{table}[hbt]
\begin{center}
    \begin{tabular}{rrcc}
    \hline
    $n$ & $\rho$ & $\rho/n$ & $B(n,\rho)$\\
    \hline
	2 & 1.4 & 0.70 & 0.28999\\
	10 & 8.0 & 0.80 &0.12166\\
	20 & 18.0 & 0.90 & 0.10921\\
	40 & 39.2 & 0.98 & 0.10544\\
     \hline
    \end{tabular}
\end{center}
    \caption{Blocking probability of an Erlang-B system as a function of the number of servers, load and system utilization.}
    \label{tab:non_constant}
\end{table}

\subsection{Heuristics}
\label{sec:qed}

The `Optimal' policy described above requires that all required parameters should be given \emph{ex ante}. Unfortunately in most settings one encounters in practice such key parameters must be inferred or forecasted based on available data, and these forecasts are bound to have a certain level of uncertainty, which the optimal policy does not account for. Below we present two heuristics that explicitly account for this uncertainty.

\subsubsection{QED Heuristics}

Congestion-related effects (delay or denial of service) are generally attributable to stochastic variability in either $\lambda$ or $\mu$. A typical rule-of-thumb -- the well known ``square-root safety staffing rule'' -- splits the amount of servers between ``base capacity'' and ``safety capacity'', with the latter being used for dealing with stochastic variability. In other words, the parameters are estimated from the statistics collected during an epoch, and for the duration of the next interval the number of servers is set to

\begin{equation}
    n = G(\rho + z_{\alpha} \sqrt{\rho})  \mbox{,}
\end{equation}

\noindent where the second term 
represents the variability hedge and takes the form of the square-root safety staffing principle, $G(\cdot)$ is defined as

\begin{equation}
G(x) = \left\{ 
\begin{array}{ll}
\lceil x \rceil & \textrm{if $P(\lceil x \rceil, \rho) \ge P(\lfloor x \rfloor, \rho)$}\\
\lfloor x \rfloor & \textrm{otherwise}
\end{array} \right. \mbox{,}
\label{eq:G}
\end{equation}

\noindent while $P(x, \rho)$ indicates the profit per unit time when $x$ servers are running and the load is $\rho$.

This kind of allocation gives rise to the Quality and Efficiency Driven (QED) regime that has been extensively studied in the literature~\cite{janssen:2011, borst:2004}.
While the amount of additional servers is proportional to $\sqrt \rho$, the decision variable $z_{\alpha}$ dictates the amount of hedging, and is a result of a ``second order'' optimization problem that seeks a suitable trade-off between the cost for servers and provided service level~\cite{borst:2004}.

Since the behavior of the square-root-rule heavily depends on the value of the parameter~$z_{\alpha}$, 
we employ an approximation that was first introduced in~\cite{mazzucco:2012} . Let $\alpha$ be the probability of all servers being busy, and assume that the
actual load is normally distributed (for large values of $\rho$ the Poisson distribution is approximately normal) with the mean value being equal to the predicted
value. Hence, by computing the quantile function (the inverse of the CDF) of the normal distribution,
$z_{\alpha}$, one can ensure that the probability of seeing all servers busy {\it does not exceed}
$\alpha$

\begin{equation}
\label{eq:z_alpha_inverse_cdf}
z_{\alpha} = \Phi^{-1}(1 - \alpha) \mbox{,}
\end{equation}

\noindent where $\Phi(\cdot)$ is the CDF of the standard normal distribution (mean 0 and variance 1).

As the reader can see, the problem reduces to find the optimal value of $\alpha$. In order to do that, we employ the approach suggested in~\cite{grassmann:1988}. At any given point in time, the system is either overloaded, and thus some jobs are being discarded, or over-provisioned, and therefore some servers are idling.
Hence, we can distinguish between two cases.

$P(n, \rho) = cn\mu - dn$ if $\rho > n$, where $n\mu$ is the maximum system throughput (the service rate, $\mu$, indicates the speed of each server). 

Similarly, if the system is over-provisioned ({\it i.e.}, $\rho \le n$) then $P(n, \rho) = c\lambda - dn$, and no job is lost.

Since the probability that the first event occurs is $\alpha$, the system is over-provisioned with probability $(1 - \alpha)$. Therefore we obtain 

\begin{equation}
P(n, \rho) = \alpha (cn\mu - dn) + (1 - \alpha) (c\lambda - dn) \mbox{.}
\end{equation}

Now, recall that the objective is to maximize the expectation of $P(n, \rho)$. In order to do that we take the first derivative of $P(n, \rho)$ with respect to $n$, which equals $-d$ for $\rho \le n$ and $c\mu - d$ for $\rho > n$

\begin{equation}
P^{\prime}(n, \rho) = \alpha (c\mu - d) + (1 - \alpha) (- d) \mbox{.}
\end{equation}

Obviously the above derivative should be set to zero in order to find the optimal number of servers. This gives the following condition for optimality

\begin{equation}
\alpha = \frac{d}{c\mu} \mbox{.}
\label{eq:alpha}
\end{equation}

Having estimated $\alpha$ one can easily compute $z_{\alpha}$ using formula~\eqref{eq:z_alpha_inverse_cdf}.

It is perhaps worth noting that depending on the relative magnitudes of $c$ and $d$, $z_{\alpha}$ might assume negative values. In that case $n < \rho$, and the system works in the so-called Efficiency-Driven regime.%
\subsubsection{Grassmann Heuristics}
\label{sec:grassmann}

The QED strategy acknowledges stochastic uncertainty by means of the square-root safety staffing rule. However it does not address the problem related to the quality of the parameters, in particular of the arrival rate. 
It is important to understand that despite the fact that sudden changes in the arrival rate are relatively small, every time a prediction is made, some error is introduced~\cite{steckley:2009}. Therefore a prediction with absolute precision is rather unlikely. %

Besides that, while servers are allocated/released at discrete points in time, the load fluctuates continuously. In other words, further errors are introduced when trying to {\it estimate} the arrival rate.
For example, when samples of the workload shown in Fig.~\ref{fig:clarknet} are averaged over a one hour interval the observed arrival rate ranges between 52.6 and 358.3 jobs/min., while it lies in the interval 15--600 jobs/min. when samples are collected every minute.
Rather than employing the sample mean, one could envisage more complex strategies. 
However experiments involving different smoothing techniques did not show any visible difference compared to the approach employing the sample mean, hence it was decided to employ the average arrival rate of the previous epoch as a input for predicting future arrivals.

\begin{figure}[ht!]
\centering
\includegraphics[width=0.45\textwidth]{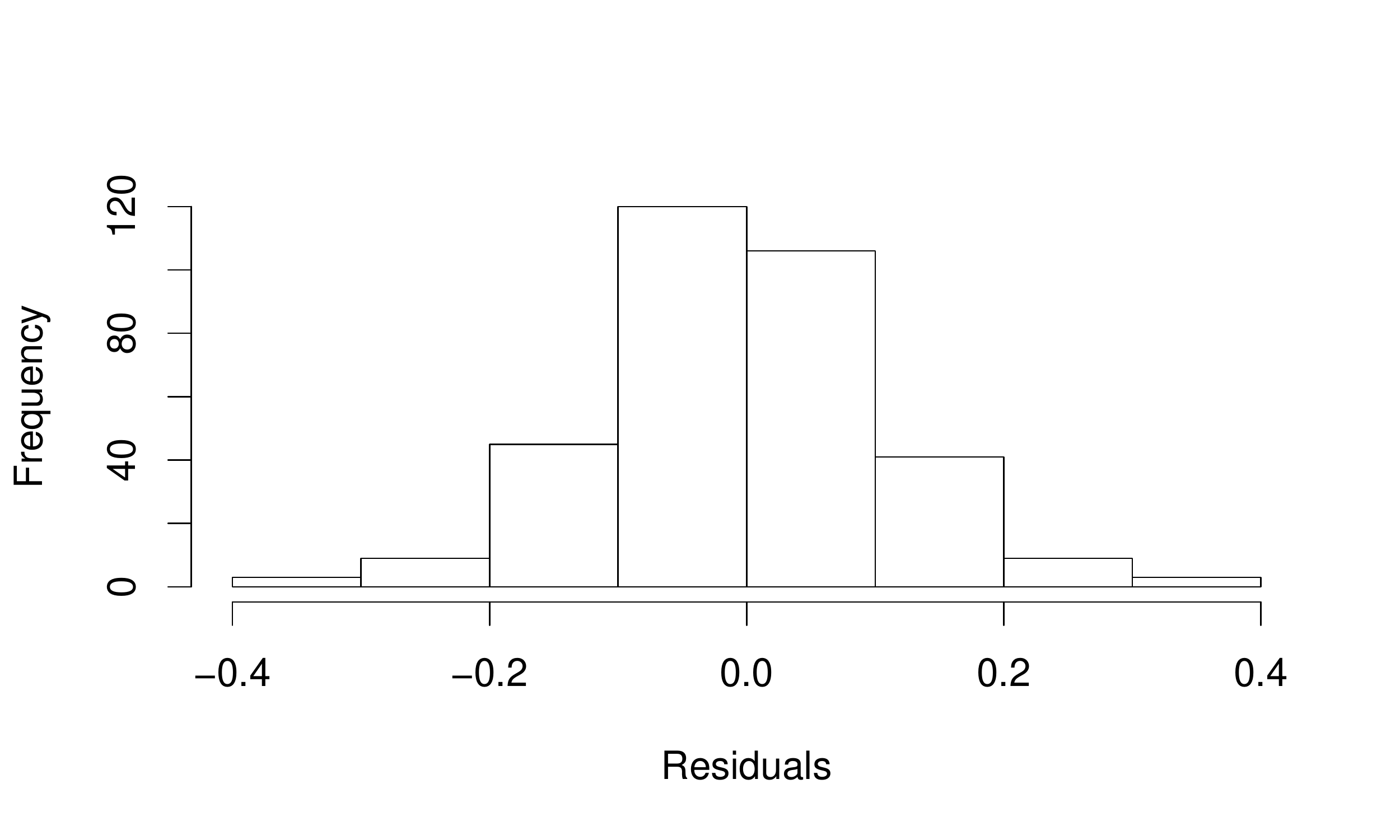}
\caption{Relative forecasting error of the forecasting algorithm applied to the ClarkNet traces shown in Figure~\ref{fig:clarknet}.}
\label{fig:hist_res}
\end{figure}

Figure~\ref{fig:hist_res} shows the distribution of the relative forecasting error we obtained when trying to predict the ClarkNet workload displayed in Figure~\ref{fig:clarknet}.
As one can see, the distribution of the forecasting error is approximately normally distributed with mean 0. 
An important implication of this finding is that about 50\% of the time the system is over-provisioned while 50\% of the time circa it is under-provisioned.

To account for forecasting errors, one can further randomize the value of $\rho$.
In this line, Grassmann~\cite{grassmann:1988} suggests to allocate servers according to

\begin{equation}
n  = G(\rho + z_{\alpha} \sqrt{\rho + Var(\rho)}) \mbox{,}
\end{equation}

\noindent where $Var(\rho)$ indicates the variance of the load, {\it i.e.}, $Var(\lambda)/\mu^{2}$ in our case, and $G(\cdot)$ is defined as in~Eq.~\eqref{eq:G}.

\section{Experimental Evaluation}
\label{sec:experiments}

In this section we present the results of a number of empirical studies we have conducted to evaluate the performance of our proposal.

\subsection{Experimental Setup}
\label{sec:experiments_setup}

Rather than employing benchmark applications such as Rubis (an auction prototype modeled after eBay), Rubbos (a bulletin board benchmark modeled after Slashdhot) or similar, we decided to test our proposal on a replica of the English edition of Wikipedia %
deployed on the Amazon Elastic EC2 cloud compute platform. Apart from the fact that the Wikipedia system arguably reflects a real large-scale deployment, the rationale behind this decision lies in the observation that while benchmarks employ synthetic workloads, deploying a replica of Wikipedia (specifically the MediaWiki application\endnote{\url{http://www.mediawiki.org/wiki/MediaWiki}}) enables us to test the system using real traces as Wikipedia snapshots and logs are publicly available. %
Finally, the model introduced in Section~\ref{sec:model} is compatible with Wikipedia's approach, insofar as Wikipedia employs denial of service to deal with spikes in demand.\endnote{\url{http://www.datacenterknowledge.com/archives/2008/06/24/a-look-inside-wikipedias-infrastructure/}}

One of the questions we faced was how to choose between the various instance types offered by Amazon EC2. Some benchmarks suggested us to employ {\tt c1.medium} instances, as they provide the best tradeoff between performance and cost.
For example while {\tt m1.small} instances can serve up to 8 jobs/sec and {\tt m1.large} instances have a throughput of 20 jobs/sec, {\tt c1.medium} can serve about 28 jobs/sec (more details in Sec.~\ref{sec:calibration}).
Thus, all servers are {\tt c1.medium} instances running in the {\tt us.east} availability zone and employing image AMI {\tt ami-e358958a} (Ubuntu Linux 11.04 32 bit, kernel~2.6.38).

Our setup consists of one node running nginx 0.8.54 to balance incoming traffic to a variable number of Apache 2.2.17 servers running MediaWiki.
Apart from nginx, the load balancer also runs a Java daemon responsible for extracting statistics from nginx, taking allocation decisions (R 2.14.0 is employed to forecast future arrivals), and updating the system configuration.
Persistent storage is provided by one MySQL v5.1.54 server, while another machine runs memcached 1.4.5. The load generator is composed of two nodes running a customized version of WikiBench~\cite{baaren:2009} on top of Oracle Java SE 1.6.0\_26, while
the roundtrip time of packets between VMs varies between 0.207 and 24.028 ms, averaging 0.579 ms.
The setup is summarized in Table~\ref{tab:setup}.

\begin{table}[hbt]
\begin{center}
    \begin{tabular}{rll}
    \hline
    {\bf Number} & {\bf Product} & {\bf Functionality}\\
    \hline
	2 & Wikibench & Load generator\\
	1 & nginx v0.8.54 & Load balancer\\
	1 & MySQL v5.1.54 & Database\\
	1 & Memcached v1.4.5 & Cache\\
	0--20 & Apache v2.2.17, & \\
	& PHP v5.3.5,& \\
	 & XCache v1.3.1 &HTTP Server\\
     \hline
    \end{tabular}
\end{center}
    \caption{Configuration.}
    \label{tab:setup}
\end{table}

\noindent {\bf Main optimizations}
We have increased the maximum number of open files (including sockets) to 20,000 (see {\tt ulimit -n)}. 
HTTP servers employ XCache 1.3.1 to cache compiled PHP code, thus preventing re-compiling the same code for every request. We have decreased the memory limit of PHP (we use PHP 5.3.5) to 32 MB in order to avoid excessive memory consumption. We have disabled logging and page visiting counters on MySQL in order not to slow down the database. Also, we have set the socket timeout to 10 seconds on both the load balancer and clients with the aim of preventing situations where a long request causes a timeout while it is being executed. %
Finally, in order to better deal with long jobs, see Table~\ref{tab:serv_times}, 
nginx was compiled with the {\tt upstream\_fair} module which routes jobs to the least-busy backend server rather than forwarding incoming requests in a round-robin manner.

\begin{table}[hbt]
\begin{center}
   \begin{tabular}{lr}
    \hline
	Percentile &  Value (ms)\\
	    \hline
	   25\%  &  71\\
	   50\% &   74\\ 
	   75\% &   79\\
	   90\% &   85\\
	   95\% &   90\\
	   99\% &  109\\ %
	   99.99\% & 9,779\\
     \hline
    \end{tabular}
\end{center}
    \caption{Service time distribution: the mean is 83~ms, while the squared coefficient of variation (variance over the square of the mean) is~8.04.}
    \label{tab:serv_times}
\end{table}

\noindent {\bf Dataset}
The database was initialized with the MediaWiki page dumps of January 15, 2011
, consisting of 166,977 articles (2.8 GB of filesystem space).
The operational dataset, however, is composed of about 1,000 articles (excluding the redirects), as we request only a portion of randomly selected articles in order to ensure that most of the requests can be served from the cache -- about 75\% of the queries are cached by MySQL while the cache hit ratio of memcached is 97\%. %
Using the whole dump would require us to use a distributed deployment for memcached, without introducing any substantial difference, apart from increasing the time required to build the cache, which is about six hours at present.
On the other hand, increasing the amount of employed data without re-provisioning the datastore would decrease the cache hit rate on both  memcached and  database, thus dramatically decreasing the throughput (if no content is cached, one server can serve only four jobs/sec).

\begin{figure}[h!tb]
\centering
\includegraphics[width=0.46\textwidth]{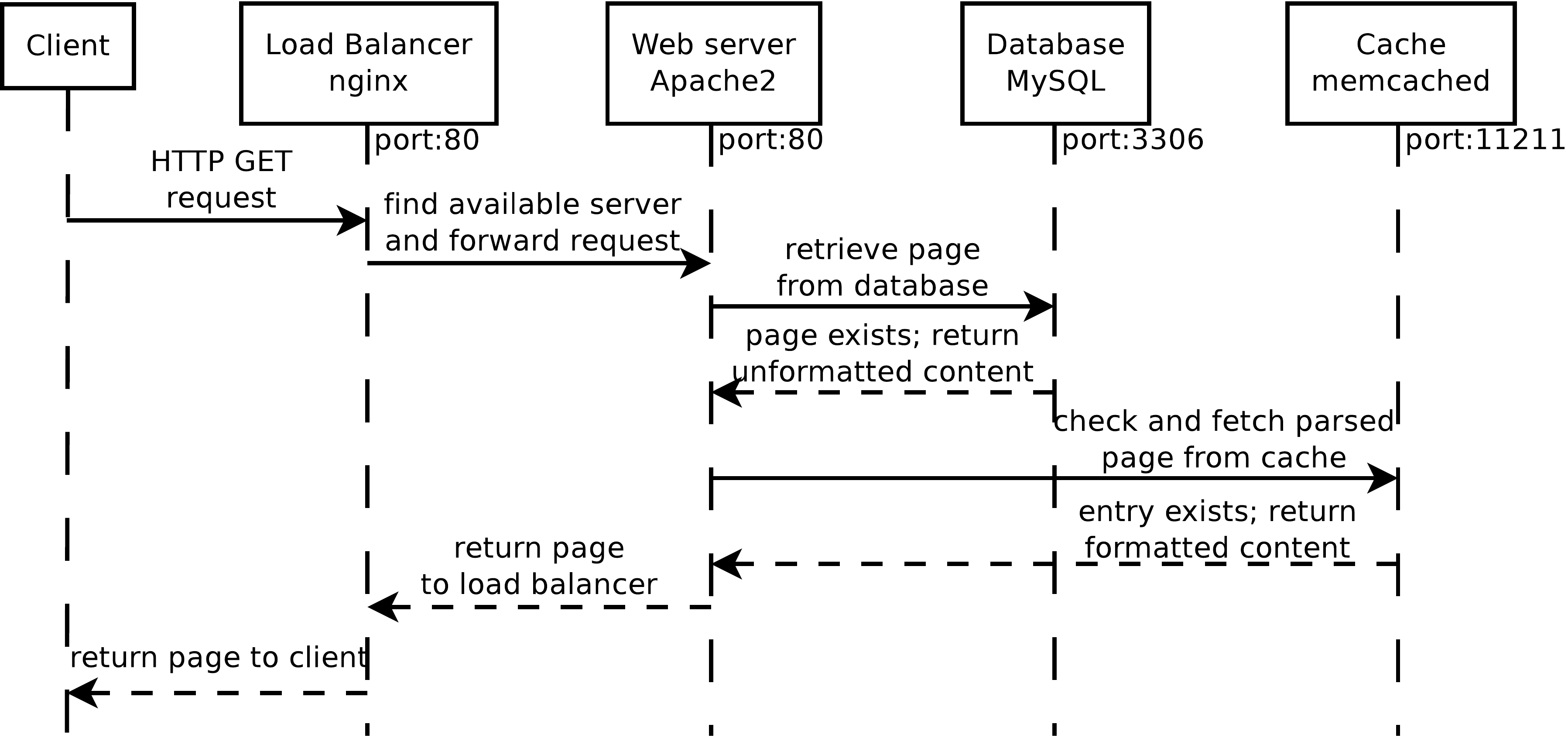}
\caption{Request execution.}
\label{fig:mediawiki_workflow}
\end{figure}

Introducing a layer providing a caching abstraction complicates the setup considerably, as jobs might be served entirely from the cache (in which case at least one database access is still required, see Figure~\ref{fig:mediawiki_workflow}), entirely from the database, or a mix of the two previous options (when the page is not in the cache, but some elements such as menus are).
Parsing the content from the database takes 1--4 seconds on a {\tt c1.medium} instance, while serving a job whose content is in the cache requires 70 ms only.
This is due to the fact that cached pages make 24.5 database queries ({\it i.e.}, number of executed statements) and 8.6 memcached accesses on average, while for non cached requests those values increase to 169.6 and 42.5 respectively.
The average file size of each request is about 64 KB.

\noindent {\bf Caveats}
A recent study of the Wikipedia workload~\cite{urdaneta:2009} found that the read to write ratio is about 480, therefore in our experiments we deal with read operations only. This caveat simplifies running the experiments considerably, as we do not have to care about ``resetting'' the database to the original state after every run, without affecting the validity of our results.
Also, while a large fraction of Wikipedia requests is due to static content requests (about 78\% of Wikipedia traffic is handled by Squid servers\endnote{\url{http://meta.wikimedia.org/wiki/Cache_strategy}}), the real problem is dealing with dynamic content. For example, in one experiment we employed four Apache servers dedicated to serve static pages only, and found that they could sustain a throughput of about 
0.5 Gbit/sec.
Therefore, we deliberately chose to deal with dynamic content only and removed Squid from our setup.

\subsection{Model Calibration}
\label{sec:calibration}

\begin{figure*}[hbt]
\centering
\subfigure[]{
\includegraphics[width=0.43\textwidth]{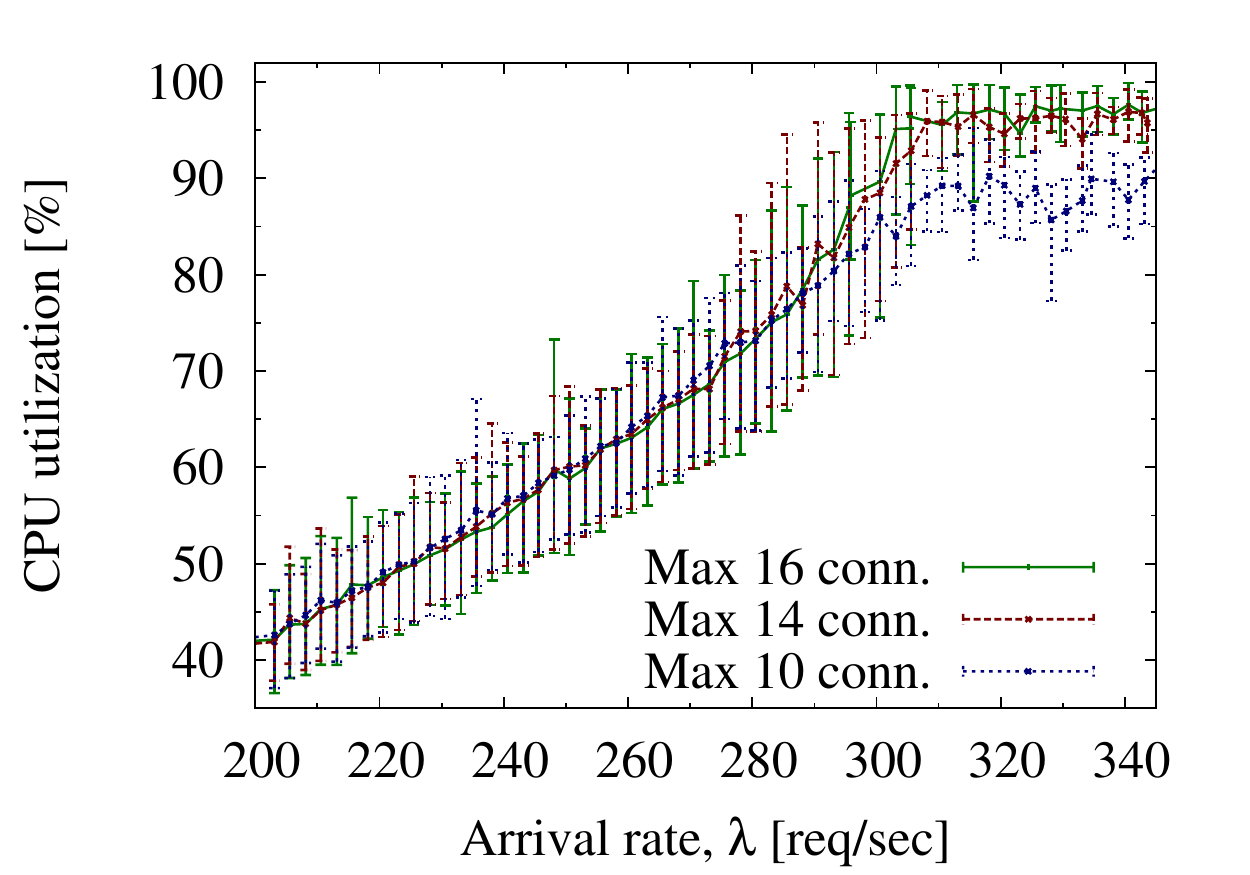}
\label{fig:c10c14c16_cpu_utilization}
}
\subfigure[]{
\includegraphics[width=0.43\textwidth]{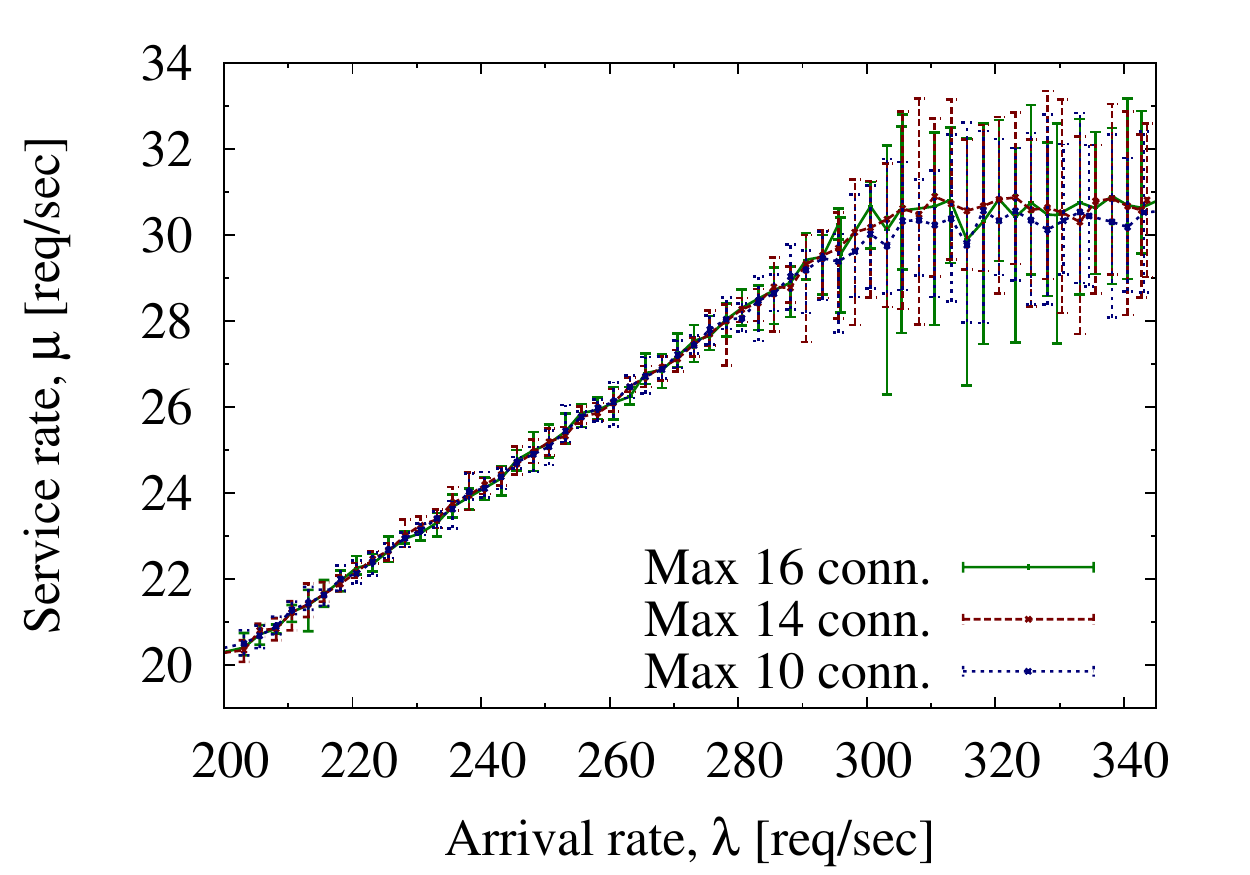}
\label{fig:c10c14c16_serv_rate}
}
\subfigure[]{
\includegraphics[width=0.43\textwidth]{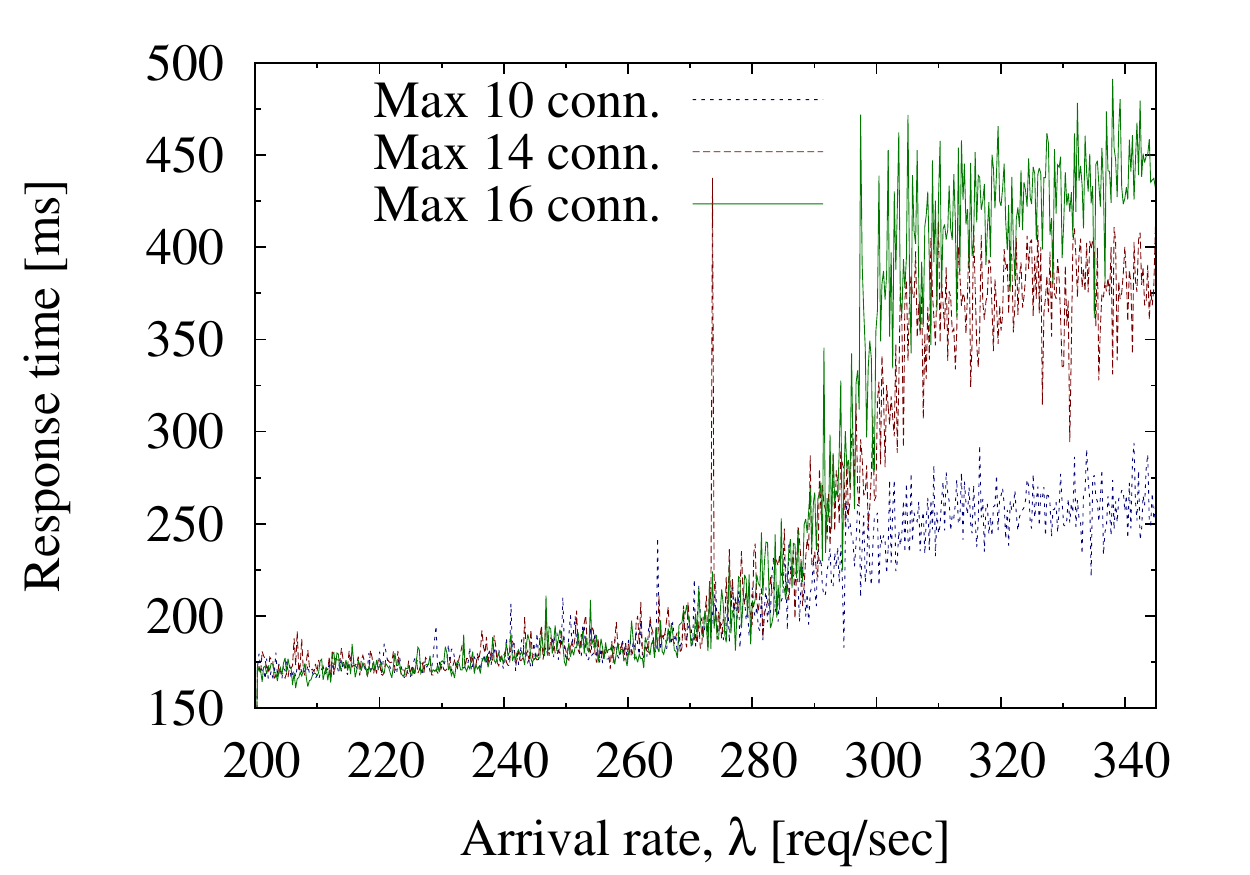}
\label{fig:c10c14c16_resp_time}
}
\subfigure{
\includegraphics[width=0.43\textwidth]{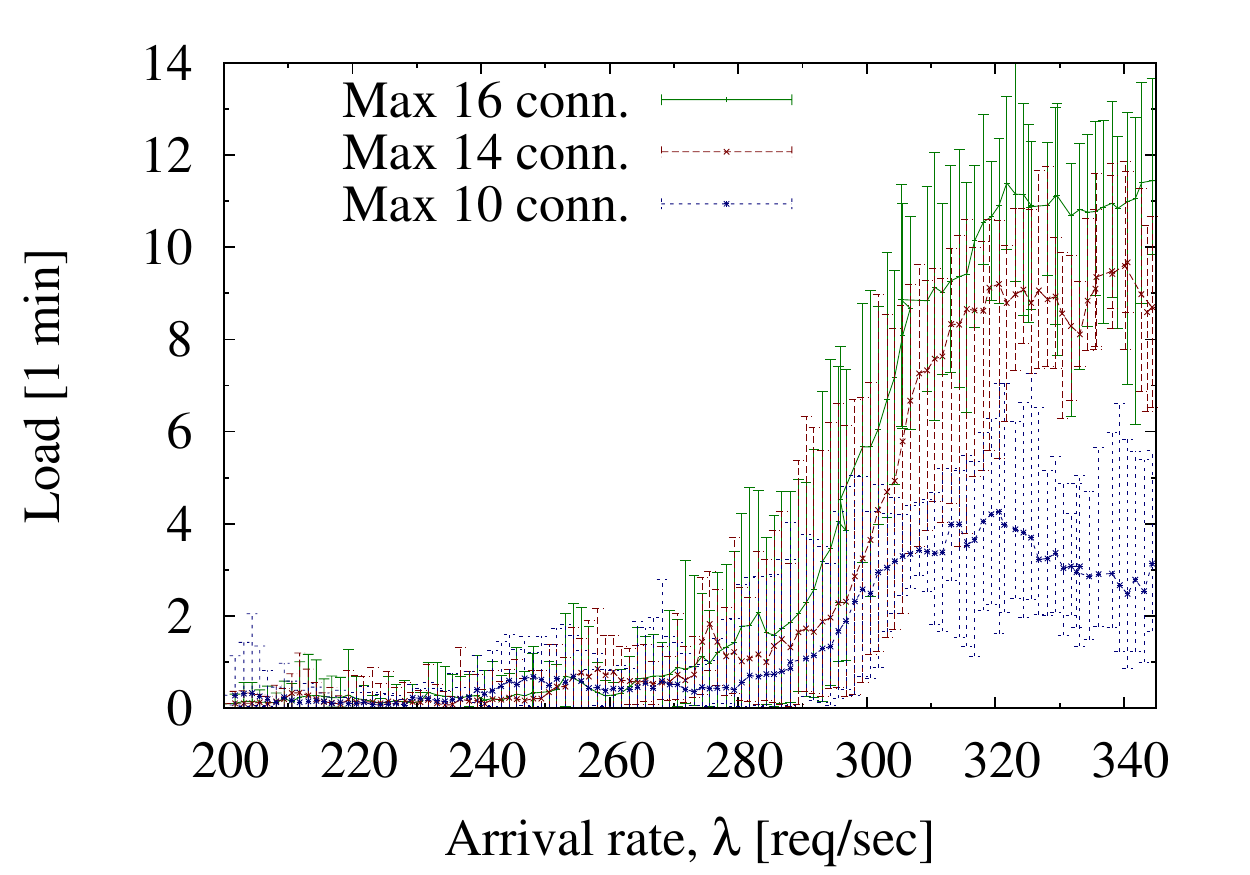}
\label{fig:c10c14c16_apache_load}
}
\caption{10 Apache servers (\texttt{c1.medium} instances) running the Wikipedia workload: (a) CPU utilization, (b)~service rate, and (c) response time and (d) Linux CPU load 
as a function of the arrival rate and maximum number of concurrent connections per server.}
\label{fig:c10c14c16}
\vspace{-3mm}
\end{figure*}

Figure~\ref{fig:c10c14c16} reports the results of an experiment we carried out with the aim of finding to which extent the value of $m$ affects the system performance. In order to do that, we benchmarked the performance of a replica of Wikipedia composed of 10 Apache application servers  subject to an increasing workload, from 200 to 345 jobs/sec over a 1-hour period, and we observed how the response time, CPU utilization and throughput changed with respect to $m$.
Since Apache employs the process-per-connection model it was easy to bound the maximum number of concurrent requests by limiting the number of connections. This limit was set on the load balancer.
Note that a similar result cannot be achieved by changing only the values of {\tt MaxClients} and {\tt ServerLimit} in Apache\endnote{\url{http://httpd.apache.org/docs/2.0/mod/mpm_common.html}}. In that case, one would lose control over the load balancing (excess jobs would wait in the TCP queue of the backend servers).
On the other hand, in the chosen setup a maximum of $n \times m$ jobs are allowed into the system, with excess traffic being silently discarded by the load balancer and not affecting future arrivals.

As it can be seen in Figure~\ref{fig:c10c14c16_cpu_utilization}, about 310~jobs/sec are necessary to saturate the system when the maximum number of concurrent connections is at least~14. At the same time the results indicate there is no visible difference in terms of throughput, see Figure~\ref{fig:c10c14c16_serv_rate}: each point represents the average speed of each server over a one minute interval, and includes the confidence interval, which was computed as the best and worst recorded throughput. As the figure shows, the variance increases with the number of concurrent connections. That might be due to a number of factors, {\it e.g.}, performance degradation, a ``noisy neighbor'' (a VM hosted on the same physical node that is using a disproportionately large part of some shared resource), or simply to the fact that the load is not spread evenly among the available machines. As an example, consider that in several occasions we noticed that one server was only able to serve 20 jobs/sec while the remaining were dealing with about 30 jobs/sec.

Finally, Figure~\ref{fig:c10c14c16_resp_time} shows the response time for different configurations, where each point represents the average over six seconds. It is worth noting that, due to resource contention, a high number of concurrent connections not only degrades the response times, but also increases their variance.
Hence, we decided to set $m = 10$, as it provides the best tradeoff between resources utilization, throughput and response time variance over a wide range of loading conditions. We have found that smaller values of $m$ would further decrease the response time variability, however the throughput would also decrease. Those experiments are not shown here in order not to clutter the charts.
Similarly, while monitoring the load %
({\tt uptime} and {\tt top} commands) we found that the Linux load increases in a super-linear manner with respect to $m$, see Figure~\ref{fig:c10c14c16_apache_load}. For example when $\lambda= 345$ jobs/sec the average one minute load increased from 2.92 ($m=10$) to 11.49 ($m=16$).
Having set the value of $m$ enables us to estimate the service rate. We have observed that {\tt c1.medium} instances are provided in a number of configurations, with the two most popular being Intel E5410 with clock rate of 2.33 GHz and 12 MB of cache an Intel E5506 running at a frequency of 2.13 GHz and equipped with 4 MB of cache (please note that even though those CPUs have four cores, only two of them are available to the guest operating system). 
This explains why the maximum achieved throughput in Fig.~\ref{fig:c10c14c16_serv_rate} varies in the range 28--33 jobs/sec circa (in other experiments we have noticed even higher values).
Hence, for what the service rate is concerned, we set $\mu = 28.571$ jobs/sec (the rationale behind considering the lower bound rather than other values is explained in Section~\ref{sec:model}).

\subsection{Results}
\label{sec:results}

Next we discuss the experiments we ran on Amazon EC2 to evaluate the policies introduced in Section~\ref{sec:policies}. For comparison reasons we measured also the performance of the `Always on' policy, a policy that uses static provisioning based on peak load, and a `Reactive' policy employing Amazon's auto-scaling feature. When the latter is employed, an additional server is launched whenever the average CPU utilization of the current servers exceeds 70\% for 15 minutes. Similarly, one server is removed every time the average CPU utilization drops below 60\% for 15 minutes.

The parameters employed by the allocation policies are summarized in Table~\ref{tab:parameters}. As for the workload, we employed a 24-hours long interval (day 10) of the ClarkNet workload, see Figure~\ref{fig:clarknet}.
Given the above parameters, the approach described in Section~\ref{sec:qed} for optimizing the amount of hedging employed by the QED and Grassmann's heuristics
suggests to employ $\alpha = 0.09722$ and $z_\alpha = 1.29754$ respectively.

\begin{table}[hbt]
\begin{center}
    \begin{tabular}{cll}
    \hline
    {\bf Parameter} & {\bf Description} & {\bf Value}\\
    \hline
    $\mu$ & Service rate & 28.571 jobs/sec\\
    $c$ & Charge per job & 0.0017\textcent\\
    $d$ & Cost per server & 17\textcent/hour\\
     $n$ & Running servers & 0--20\\
     $m$ & Max. concurrent & \\
    & connections per & 10\\&server&\\
     \hline
    \end{tabular}
\end{center}
    \caption{Parameters.}
    \label{tab:parameters}
\end{table}

\begin{figure}[b!]
\centering
\includegraphics[width=0.48\textwidth]{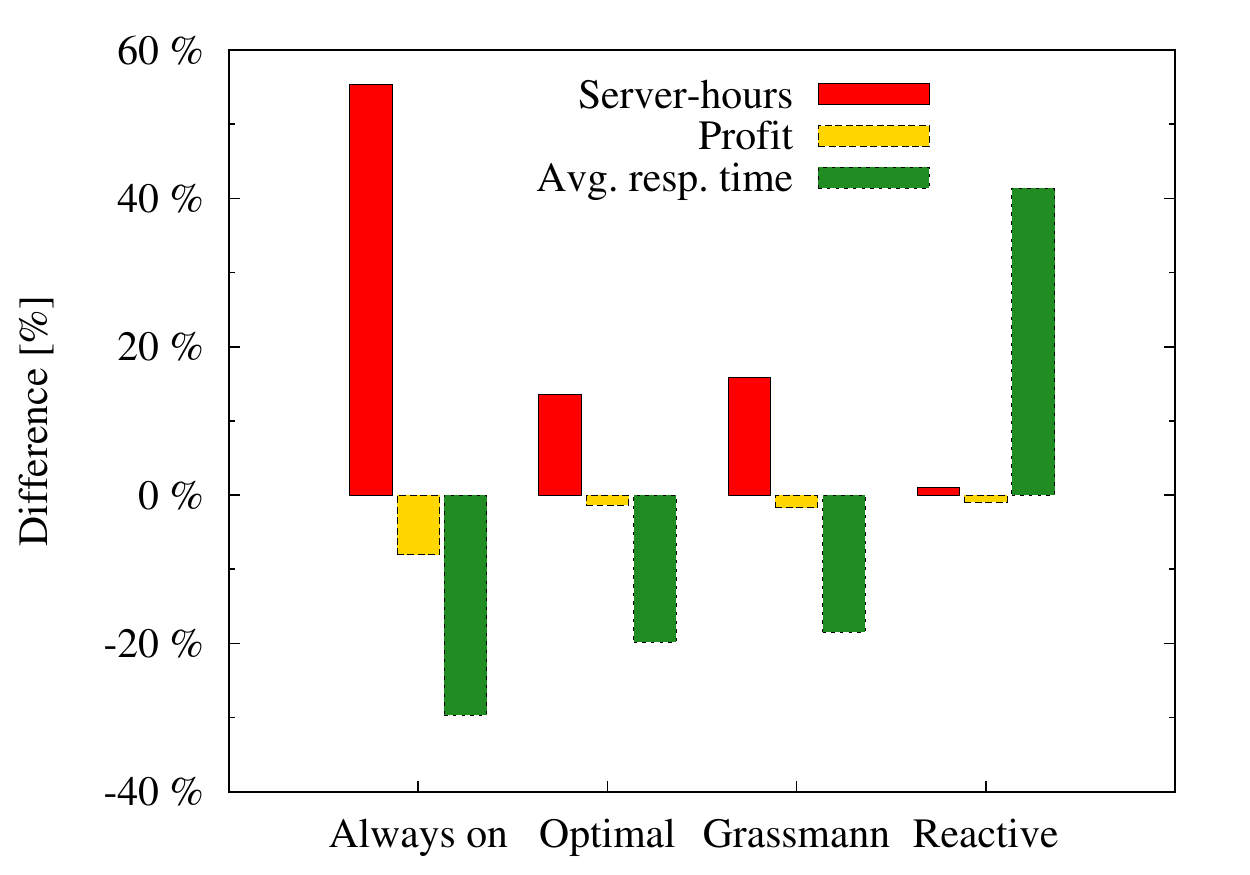}
\caption{Performance of the heuristics compared to that obtained by the `QED' policy.} %
\label{fig:comparison}
\end{figure}

\begin{figure}[b!]
\centering
\includegraphics[width=0.48\textwidth]{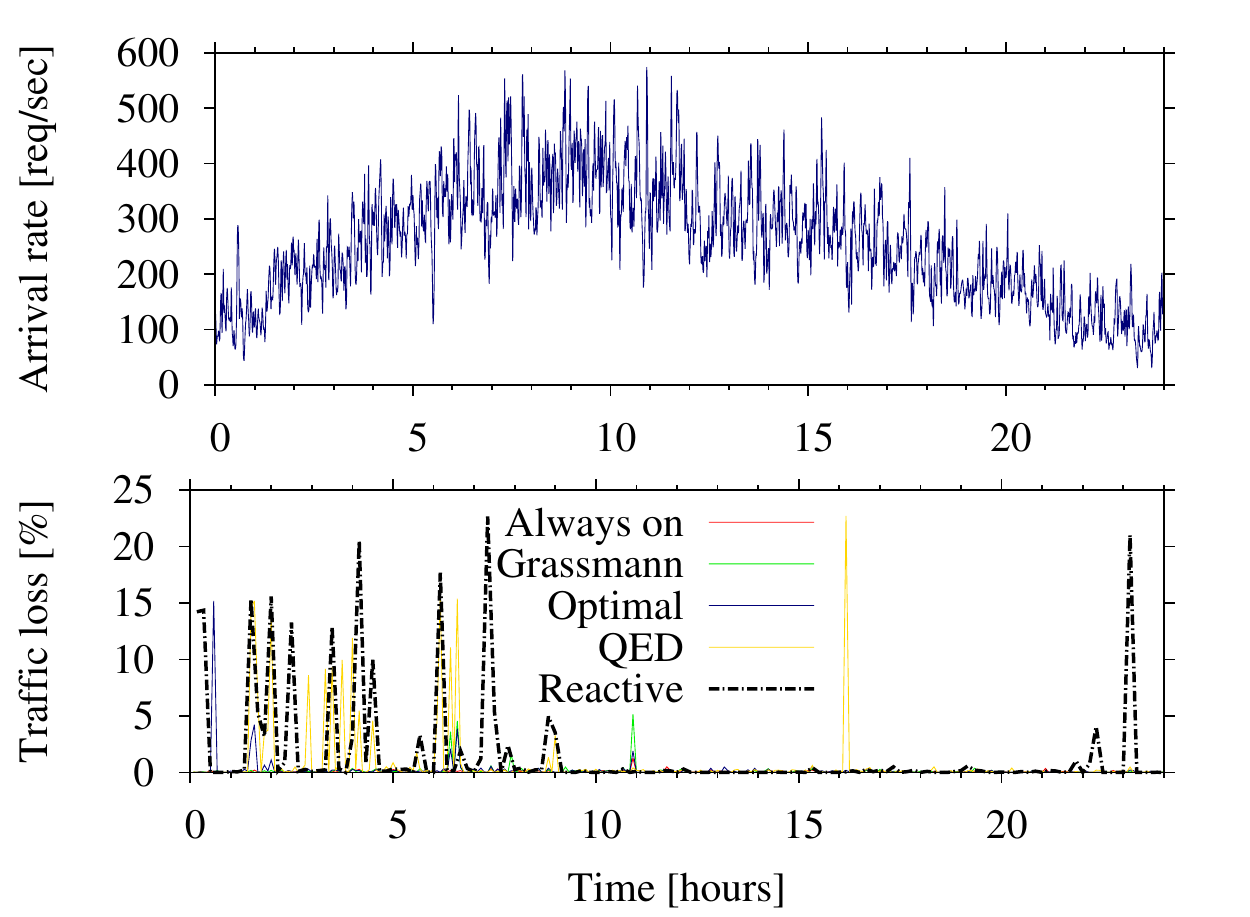}
\caption{Arrival rate and traffic lost as a percentage of the arrival rate for different policies. Every point represents the average over a 30 seconds interval. The spike in the traffic loss of the `Optimal' policy during the first hour is due to a temporary increase in the latency of MySQL: the round-trip time jumped from 0.3 to over 12 ms, however no change in the CPU utilization was observed.}
\label{fig:errors}
\end{figure}

Fig.~\ref{fig:comparison} compares the achieved profit over the 24 hours period, the number of server-hours employed as well as the average response time.
During that period about 22.2 million page requests arrived into the system; the number of accepted jobs was slightly smaller, and depends on the employed policy, see Table~\ref{tab:summary}. Servers allocation occurred every hour according to the load predicted by the algorithm described in Section~\ref{sec:param_estimation}, while the arrival rate changed every minute (see top part of Fig.~\ref{fig:errors}). Hence, the assumptions made by the `Optimal' policy, namely known parameters and stationary traffic, are violated.

\noindent {\bf Always on}
As one can see, all dynamic policies outperform the `Always on' heuristic. Given the static nature of that approach, it naturally leads to the highest number of server-hours. Due to the level of over-provisioning, the profit is the lowest but on the other hand the response times and number of jobs lost are also the lowest.

\noindent {\bf N.B.} It should be noted that the profit earned by the `Always on' heuristic is only 6.6\% smaller than that of the `Optimal' because, in order to discourage denial of service, we deliberately chose a large value of $c$ compared to $d$ (see Table~\ref{tab:parameters} and discussion in Sec.~\ref{sec:model}). If the charge was smaller, the profit earned by the `Always on' policy would have been lower (possibly negative).
Also, the performance of the static policy heavily depends on the average system utilization, which in this scenario was about 50\% over the 24-hours period, as well as on its peak-to-average ratio.

\begin{figure}[t!]
\centering
\includegraphics[width=0.48\textwidth]{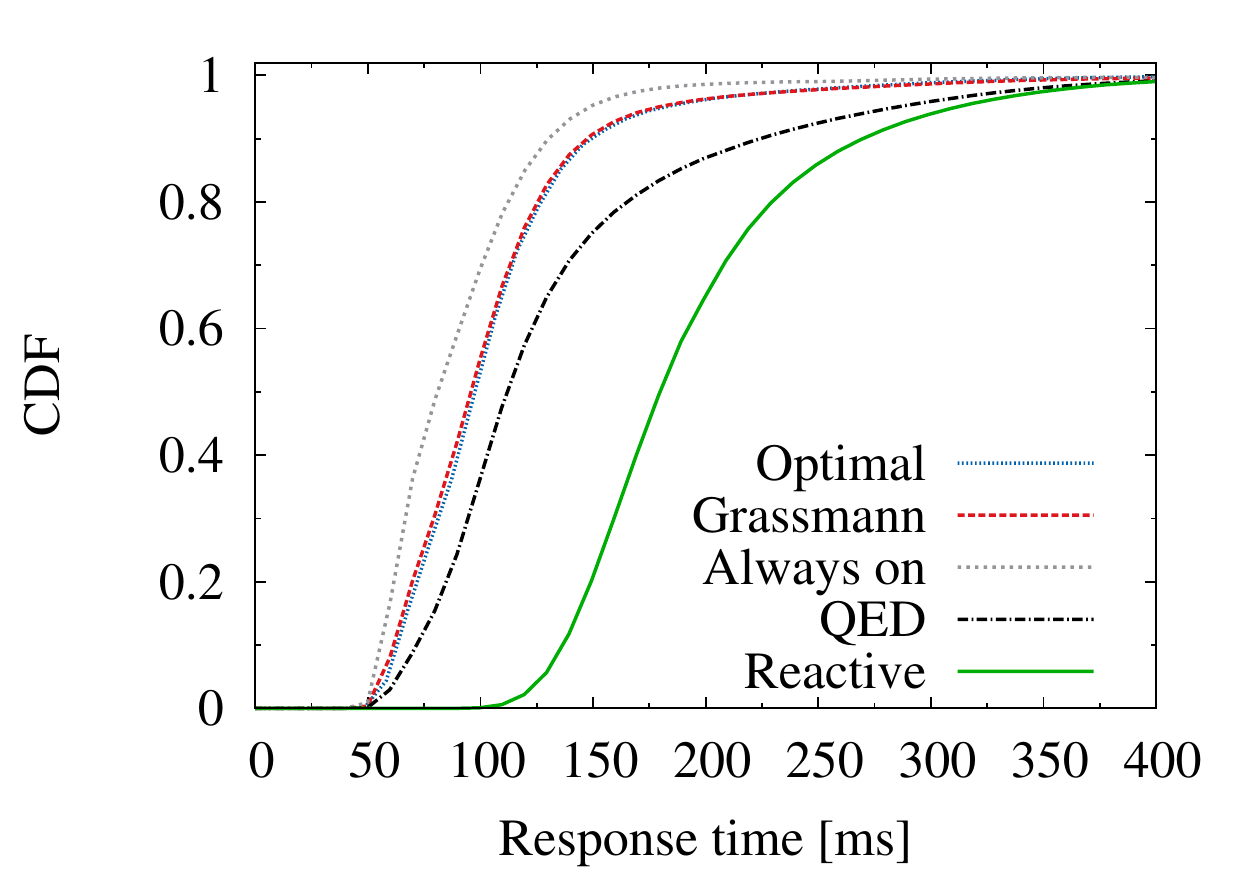}
\caption{Cumulative distribution function (CDF) of the response times for different policies.}
\label{fig:cdf}
\end{figure}

\noindent {\bf Dynamic policies} Dynamic provision enables the system to better adapt to incoming user demand, thus employing a smaller number of server-hours. For what the profit is concerned, the `QED' heuristic seems to provide the best configuration, however the difference compared to the other two algorithms is really small (less than 5\%). However, due to the fact that the `QED' heuristic uses the smallest number of server-hours, it exhibits the highest response times. Also, an adverse effect of QED is that it frequently under-provisions. This is evident in Figure~\ref{fig:errors} and Table~\ref{tab:summary} which shows the job loss per policy.
Grassmann's heuristic is close to the optimal policy in all three parameters. The `Optimal' policy slightly outperforms Grassmann's in terms of profit, while the achieved response times are about the same (Figure~\ref{fig:cdf}).
Table~\ref{tab:summary} shows that the `Optimal' policy also slightly outperforms Grassmann's in terms of job loss.
Finally, the `Reactive' policy employs about the same amount of server-hours as the `QED' policy, thus leading to a similar profit.
However, due to the fact that it re-provisions in a reactive manner, it adds servers only after servers become overloaded; therefore it shows unacceptable results for what the other metrics are concerned:
the response time is on average 42\% higher than that of the `QED' heuristic and about twice that of the `Optimal' policy, while the job loss is about one order of magnitude higher than that of the `Optimal'.

Regarding the overhead of the resource allocation policies, all algorithms (including the optimal one) can easily find the solution for more than 100,000 servers in less than a second. In fact, the execution time of the policy is dominated by the execution time of the forecasting algorithm -- which is needed by all dynamic policies.

\begin{table*}[b!th]
\begin{center}
    \begin{tabular}{lccrrcr}
    \toprule
    & &  \multicolumn{2}{c}{Jobs} & & \multicolumn{2}{c}{Response time}\\
        \cmidrule(r){3-4}
        \cmidrule(r){6-7}
    &	Server hours & Arrived & Lost & Profit (\$) & Avg. (ms) & $cv^{2}$\\
        \midrule
    {\bf Optimal} & 351 & 22,172,169 & 32,290 & 316.7 & 116.0 & 0.295\\
    {\bf Always on} & 480 & 22,171,947 & 2,149 & 295.3 & 101.8 & 0.929\\
    {\bf QED} & 309 & 22,164,522 & 184,622 & 321.1 & 144.7 & 0.619\\
    {\bf Grassmann} & 358 & 22,172,164 & 28,656 & 315.6 & 118.0 & 1.519\\
    {\bf Reactive} & 312 & 22,169,548 & 343,763 & 318.0 & 204.6 & 0.152\\ 
    \bottomrule
    \end{tabular}
    \end{center}
        \caption{Summary of the results.}
    \label{tab:summary}
\end{table*}

An attentive reader might note that, while the `QED'  heuristics was tuned for a job loss of about 10\%  (specifically, ${\alpha} = 0.09722$), the actual job loss was about one order of magnitude smaller (cf. Table~\ref{tab:summary}). %
This behavior can be explained by the fact that Equation~\eqref{eq:alpha} assumes that the traffic is normally distributed and it does not take into account the accuracy of the arrival rate prediction. Yet, the prediction accuracy can vary greatly depending on the method chosen and the type of workload. For example, while the average error of the modified Holt-Winter's algorithm described in Section~\ref{sec:param_estimation} is about 8\% when applied to the considered workload, that of the Double Exponential Smoothing is about 16\%.
And since the prediction accuracy is not taken into account, it is more likely that the system is over-provisioned rather than over-loaded, which implies that the observed ${\alpha}$ is lower than the one given as parameter to heuristic~\eqref{eq:alpha}.

\section{Related Work}
\label{sec:relatedwork}

Cluster sizing related issues are not unique to cloud platforms. However, what makes the cloud unique in this respect is that cluster
sizing problems arise much more frequently here due to the pay-as-you-go and elastic nature. Furthermore, SaaS providers have little to no control over the underlying infrastructure of public clouds. Private clouds, on the other hand, are in most cases nothing more than clusters with a virtualization layer, but remain under the control of the organization.

While several resource allocation policies have been proposed in the literature, %
apart from some rare exception ({\it e.g.}, see~\cite{hunter:2011}) previous proposals have not been tested on the cloud, they do not explicitly acknowledge errors related to parameters estimation/forecasting %
nor do they consider that acquire and release operations should be performed at discrete points only (this is especially true for control theory based approaches).

The most closely related work can perhaps be found in~\cite{lim:2010} and~\cite{gulati:2011}, which present strategies aimed at allocating elastic storage nodes with the aim of delivering acceptable service levels. The former is based on control theory, so it simply adapts to observed performance and/or conditions change, while the latter monitors several metrics in order to constructs and adapts approximate black-box performance models of storage devices automatically, aiming at linking device throughput and latency to outstanding IOs.
Chen {\it et al.}~\cite{chen:2005} introduced a queuing model for controlling the energy consumption of service provisioning systems subject to Service Level Agreements. The authors, however, do not acknowledge the time and energy wasted during state changes, while we have shown that the time required to add new servers can play an important role. 
Ardagna {\it et al.} \cite{ardagna:2010} discuss the resource allocation problem
in multi-tier virtualized systems with the goal of meeting the QoS requirements
while minimizing energy costs. 
However it can not be employed on large scale deployments, as the problem is NP-hard. Furthermore the model assumes a closed queueing network, which is not very suitable for Internet deployments where the potential number of users is very large, thus under-estimating the number of required resources (for a given loading scenario, the performance of closed systems is much better than that of open systems~\cite{schroeder:2006a}).
{Chase {\it et al.}~\cite{chase:2001} presented an architecture for resource management of server farms. There the goal is to reduce energy consumption, while the SLAs are assumed to be flexible ({\it i.e.}, service degradation is a viable option), while the services ``bid'' for resources as a function of delivered performance.
Finally, Hu {\it et al}~\cite{hu:2009} investigate how to deliver response time guarantees in a multi-server and multi-class setting hosted on the Cloud by means of allocation policies only. 

\section{Conclusion and Future Work}
\label{sec:conclusion}

The main contribution of the paper is a comparative evaluation of three policies for addressing the problem faced by a SaaS provider aiming at maximizing its profit while delivering a service on a pay-per-transaction basis, using resources provided by an IaaS provider who bills per server-hour. Experimental results showed that all three policies clearly outperform an `Always on' policy where a fixed number of servers are kept in use as well as a `Reactive' approach based on Amazon's auto scaling mechanism. Also, the policy based on the optimization of the SaaS provider's utility function slightly outperforms the QED and Grassmann heuristics. A secondary contribution of the paper is an approach to implement the optimal policy so that it scales up to thousands of servers.

While the experimental results are encouraging and demonstrate the potential benefits of dynamic profit maximization policies on the cloud, there is room for improvement in the implementation and calibration of the studied policies. Regarding the `Optimal' policy, we opted to approximate the number of servers in the $G/GI/n/n$ model as the number of running virtual servers, thus treating each virtual server as a ``black box''. One might argue that a better approach would be to use the total number of cores instead. It turns out that this alternative is still a poor approximation. Instead, according to some experiments we have run, employing the total number of connections ({\it i.e.}, $n \times m$) as the number of ``servers'' and taking the service rate as $\mu / m$ improves the quality of the approximation to some extent, though in practice this approach requires one to carefully select the number of connections $m$ and this calibration needs to be done in a system-specific manner. A possible avenue for future work is to extend the policies with a method to tune parameter $m$ and then optimize the total number of connections supported by the allocated virtual servers, as opposed to optimizing the number of virtual servers. The idea of optimizing based on the total number of connections could also allow us to take into account performance differences across servers ({\it i.e.}, different servers might support different number of connections).

The model described in Section~\ref{sec:model} focuses on the application layer and does not explicitly acknowledge the database layer. A finer-grained model that supports dynamic scaling of the database layer might improve the performance of the policy and make it applicable to applications with lower read-to-write ratios than Wikipedia. Dynamically scaling up the database layer brings in additional problems, such as the high delays associated with powering on/off database servers and data replication and partitioning problems, which deserve a separate treatment.

\vspace{-1mm}

\section*{Acknowledgements}
This work was partly funded by ERDF via the Estonian Competence Centre Programme and by the European Commission via the REMICS project (FP7-257793).

\theendnotes

\footnotesize \bibliographystyle{IEEEtran}
\bibliography{green-computing}

\end{document}